# High-efficiency superconducting nanowire single-photon detectors fabricated from MoSi thin-films

V. B. Verma[1], B. Korzh[2], F. Bussières[2], R. D. Horansky[1], S. D. Dyer[1], A. E. Lita[1], I. Vayshenker[1], F. Marsili[3], M. D. Shaw[3], H. Zbinden[2], R. P. Mirin[1], and S. W. Nam[1]

[1]*National Institute of Standards and Technology, 325 Broadway, Boulder, CO 80305, USA*

[2]*Group of Applied Physics, University of Geneva, CH-1211 Geneva 4, Switzerland*

[3]*Jet Propulsion Laboratory, California Institute of Technology, 4800 Oak Grove Dr., Pasadena, California 91109, USA*

We demonstrate high-efficiency superconducting nanowire single-photon detectors (SNSPDs) fabricated from MoSi thin-films. We measure a maximum system detection efficiency (SDE) of 87 ± 0.5 % at 1542 nm at a temperature of 0.7 K, with a jitter of 76 ps, maximum count rate approaching 10 MHz, and polarization dependence as low as 3.4 ± 0.7 % The SDE curves show saturation of the internal efficiency similar to WSi-based SNSPDs at temperatures as high as 2.3 K. We show that at similar cryogenic temperatures, MoSi SNSPDs achieve efficiencies comparable to WSi-based SNSPDs with nearly a factor of two reduction in jitter.

Recent progress in the development of superconducting nanowire single-photon detectors (SNSPDs) has been rapid. To a large extent, this progress has resulted from materials research and development. Improvements in the deposition techniques for polycrystalline superconductors such as NbN and NbTiN SNSPDs have led to system detection efficiencies (SDE) above 70 % at a wavelength of 1550 nm, representing nearly an order of magnitude improvement over the first SNSPDs fabricated from NbN.[1,2] SNSPDs based on the amorphous superconductor $W_xSi_{1-x}$ have demonstrated better than 90% SDE.[3] In addition to their high efficiencies, SNSPDs have also demonstrated fast recovery times (< 10 ns),[4-7] low jitter (< 100 ps),[8] and low intrinsic dark count rates (< 1 count per second, cps).[3] They are ideal detectors for experiments in quantum optics,[9-12] as well as applications such as long-distance ground-to-space optical communications,[13] characterization of single-photon sources and photon pair sources,[14,15] light detection and ranging (LIDAR),[16] distributed fiber sensing,[17] and integrated circuit testing.[18]

Amorphous materials such as WSi are particularly desirable for the fabrication of SNSPDs due to their high degree of homogeneity and uniformity over large areas. Due to the lack of a well-defined crystal structure, amorphous superconductors can be deposited on virtually any substrate without significant degradation in material properties.[19] Thus, SNSPDs fabricated from these materials can easily be embedded inside of a dielectric optical stack to enhance absorption at a particular wavelength. Although high efficiencies have been reported with WSi at temperatures as high as 2.5 K, the jitter is also high (~150-200 ps) relative to SNSPDs fabricated from NbN or NbTiN due to lower switching currents.[20] The high efficiencies reported with WSi have instigated research into other amorphous superconductors with higher $T_C$ such as MoGe[21] and MoSi[22] which may allow operation at higher temperatures with lower jitter, thus simplifying and reducing the cost of the cryogenics. However, the efficiencies reported with these materials have so far been well below the efficiencies reported with WSi due to the lack of an optical cavity designed to enhance absorption.[3,21,22]

Here we report on the performance of MoSi SNSPDs embedded inside of an optical stack designed to enhance absorption at a wavelength of 1550 nm. We demonstrate efficiencies of 87 ± 0.5 % at 1542 nm, approaching the best efficiency achieved to date with WSi-based devices (93% at 1550 nm)[3], with significantly lower jitter (76 ps at 0.7 K) than WSi (~150 ps at 0.12 K).[3]

The fabrication process begins with the deposition of gold mirrors on a 3 inch Si wafer by electron beam evaporation and liftoff. The mirrors consist of 80 nm of gold with a 2 nm Ti adhesion layer below and above the mirror. A quarter-wave spacer layer consisting of 235 nm of $SiO_2$ is then deposited by plasma-enhanced chemical vapor deposition (PECVD), and Ti/Au contact pads are patterned by liftoff. The 6.6 nm-thick $Mo_{0.8}Si_{0.2}$ film is deposited by DC magnetron sputtering from an alloy target at room temperature and capped with 2 nm of amorphous Si to prevent oxidation. After etching a 20 µm-wide strip between the gold contact pads, electron beam lithography and etching in an $SF_6$ plasma are used to define the nanowire meanders. An antireflection coating is deposited on the top surface consisting of 294 nm $SiO_2$, 234 nm $SiN_x$, 169 nm $SiO_2$, and 161 nm $SiN_x$. A keyhole shape is then etched through the Si wafer around each SNSPD, which can then be removed from the wafer and self-aligned to a single-mode optical fiber to within ± 3 µm.[23] Each nanowire meander covers an active area of 16 µm × 16 µm which is larger than the 10 µm mode-field diameter of a standard single-mode fiber to allow for slight misalignment.

Here we show measurements of two SNSPDs from the same wafer. One SNSPD was characterized at a temperature of 0.7 K using a water-cooled closed-cycle cryocooler augmented by a pumped LHe4 unit, while the other was characterized in an air-cooled Gifford-McMahon cryocooler with 100 mW of cooling power operating at 2.3 K.

A 1542 nm CW laser attenuated to a mean photon number of ~ 300,000 photons/s was used for the measurements at 0.7 K. The calibration of the input power to the SNSPD was performed as outlined in Ref. 3 using a NIST-calibrated InGaAs power meter. Although the detector was optimized for a wavelength of 1550 nm, measurements were performed at 1542 nm because the power meter was calibrated at that wavelength. The efficiency at 1542 nm was modelled to be 98.7%, only 0.4% lower than the maximum at 1550 nm. Fig. 1 shows the system detection efficiency (SDE) and background count rate (BCR) as a function of bias current ($I_B$) normalized to the switching current ($I_{SW}$=9.5 µA, the bias current at which the device switches from the superconducting to the normal state) for a nanowire with a nominal width of 130 nm and pitch (center-to-center spacing between nanowires in the meander) of 215 nm. The SDE was measured after both maximizing (SDEmax) and minimizing (SDEmin) the detector counts as a function of the polarization of the incident light. The maximum and minimum SDE were determined from a 6-state Stokes measurement of count rates from the detector. At $I_B$ ~ 0.9$I_{SW}$, the maximum SDE was 87 ± 0.5 % and the minimum SDE was 84 ± 0.4 % with a BCR of ~ 70 cps. Note the low polarization dependence of only 3.4 ± 0.7 %, which we attribute to the high fill factor of the device and design of the optical stack. The polarization dependence is significantly lower than most results reported to date with both NbN[24] and WSi-based[3] SNSPDs. The BCR was reduced by coiling the optical fiber close to the detector in order to increase the loss for long-wavelength blackbody radiation which is the dominant contribution to the BCR well below the switching current. The fiber was coiled 5.5 times with a 35 mm diameter. Loss due to the coil was measured to be ~0.1 dB.

The optical stack for these SNSPDs was designed for maximum absorption (99.1%) at a wavelength of 1550 nm. In addition to loss introduced by the fiber coil, discrepancy between the simulated absorption and measured SDE could be attributed to a number of factors including misalignment between the fiber and device area, differences between the simulated and experimentally obtained optical constants and thicknesses of the layers of the optical stack, and an intrinsic material efficiency of converting absorbed photons to output pulses that is less than unity. Further research is required in order to understand the dominant contribution to the discrepancy between experiment and theory.

Figure 2 shows an output pulse from the SNSPD obtained using a chain of a 500 MHz amplifier with 28 dB of gain followed by a 1 GHz amplifier with 23 dB of gain, both operated at room temperature. The 1/$e$ decay time of the pulse is 35 ns. From this value of the decay time and the geometry of the nanowire, we extract a kinetic inductance of 190 pH/square for the 6.6 nm-thick MoSi film. This value is somewhat lower than that reported for typical WSi films, which can range between 200-350 pH/square depending on the composition and thickness of the film. The jitter of the detector system ($J_S$) was measured using a 1550 nm picosecond pulsed laser and 1 GHz oscilloscope to register the inter-arrival time between the laser pulse and the SNSPD pulse. The inset to Fig. 2 shows the instrument response function (IRF) of the detector system biased at $I_B = 0.95 I_{SW}$, which is the bias current corresponding to maximum SDE and a BCR of ~ 100 cps. We estimated the system jitter to be the full width at half maximum (FWHM) of the IRF, $J_s$ = 76 ps.

Measurements were also performed in a different cryogenic system operating at 2.3 K and with a narrower nanowire device (110 nm with a pitch of 185 nm). The cryogenic system is an air-cooled Gifford-McMahon cryocooler with 100 mW of cooling power. Fig. 3 shows the

SDE and BCR as a function of $I_B$ normalized to the switching current $I_{SW}$=4.1 µA. Measurements were performed with a CW 1550 nm laser with a photon flux of ~ 100,000 photons/s. The switching current is lower than the device described above ($I_{SW}$=9.5 µA) due to the use of a narrower nanowire combined with the higher operating temperature. Maximum SDE for this device is 79 ± 2 % with the polarization optimized to maximize the count rate. The BCR was below 10 cps for $I_B < 0.9I_{SW}$. A coil was also used in this measurement to reduce background counts due to blackbody photons in the fiber. Despite the increased temperature, we observed saturation of the internal efficiency similar to lower temperatures.

In order to reduce jitter and obtain the maximum possible count rate from the detector, a cryogenic preamplifier was installed at the 40 K stage of the Gifford-McMahon cryocooler with 28 dB of gain, a bandwidth of 1 GHz, and a noise figure of 1.5 dB at 1575 MHz. The resulting signal was further amplified by a room-temperature amplifier with 33 dB of gain and bandwidth of 1 GHz. A 600 MHz filter was used following the output of the room-temperature amplifier to reduce noise on the output pulse. Figure 4 shows a voltage pulse from the SNSPD. The jitter of the detector system was measured using a 1550 nm pulsed laser with a FWHM of ~ 33 ps and a time-correlated single-photon counting card to register the inter-arrival time between the laser pulse and SNSPD pulse. The inset to Fig. 4 shows the instrument response function (IRF) of the detector system biased at $I_B = 0.95I_{SW}$, which is the bias current corresponding to an SDE of 78 % and a BCR of ~ 100 cps. The system jitter is $J_s$= 99 ps. Figure 5 shows the system jitter as a function of normalized bias current. Figure 6 shows a plot of the count rate as a function of the input photon flux. The count rate remains linear up to photon fluxes approaching $10^7$ photons/s.

In conclusion, we have demonstrated high-efficiency MoSi SNSPDs embedded in a dielectric stack to enhance absorption. The SNSPDs show a saturated internal efficiency as high as 87 % at a wavelength of 1542 nm with dark count rates below 100 cps.  The lowest measured jitter was 76 ps at a temperature of 0.7 K. We demonstrated that even at a temperature of 2.3 K the MoSi SNSPDs exhibit high saturated internal efficiency. The devices remain linear in count rate with input photon fluxes approaching $10^7$ photons/s, and show a low polarization dependence of only 3.4%.

We acknowledge Claudio Barreiro for useful discussions, and the Swiss Federal Institute of Metrology (METAS) for the calibration of the power meters.

**Figures**

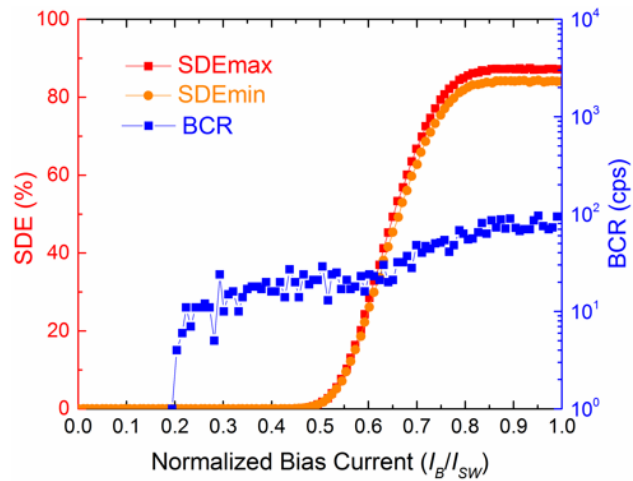

Fig. 1 System detection efficiency (SDE) and background count rate (BCR) as a function of normalized bias current. The switching current of the detector ($I_{SW}$) was 9.5 µA.

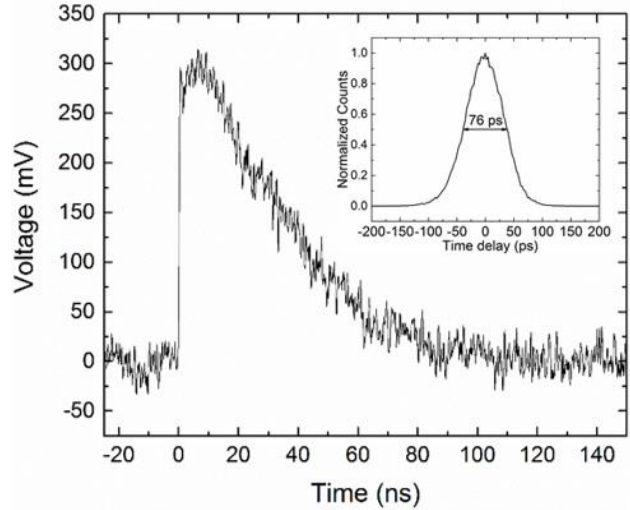

Fig. 2 Voltage pulse from the SNSPD obtained using two room temperature amplifiers with 51 dB total gain. Inset shows the instrument response function (IRF) of the detector system. The jitter, defined as the full width at half maximum (FWHM) of the IRF, is 76 ps.

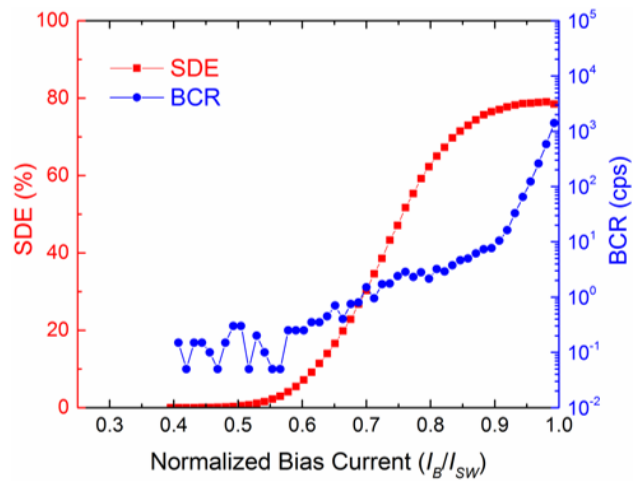

Fig. 3 System detection efficiency (SDE) and background count rate (BCR) as a function of normalized bias current for a 110 nm-wide nanowire operated at 2.4 K. The switching current of the detector ($I_{SW}$) was 4.1 µA.

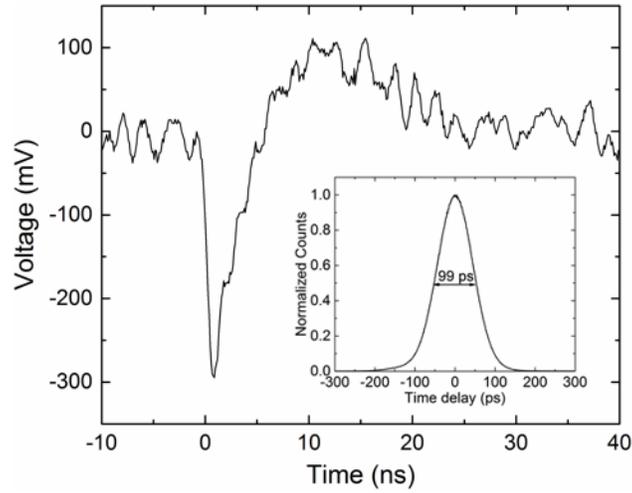

Fig. 4 Voltage pulse from the 110 nm-wide SNSPD at 2.4 K obtained using a cryogenic amplifier at 40 K and a room temperature amplifier. Inset shows the instrument response function (IRF) of the detector system. The system jitter is 99 ps.

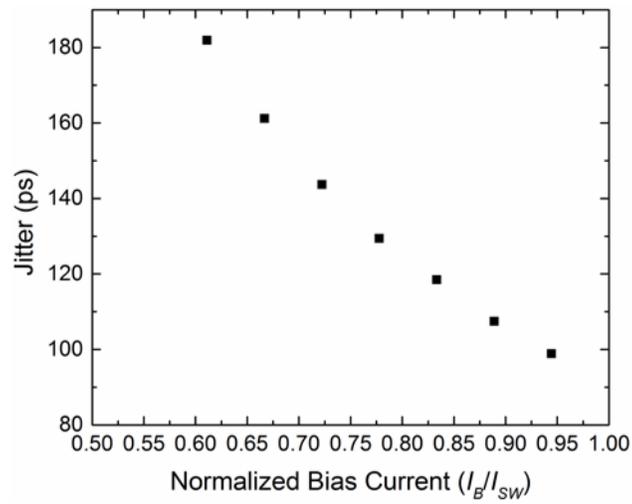

Fig. 5 Experimentally measured system jitter as a function of normalized bias current.

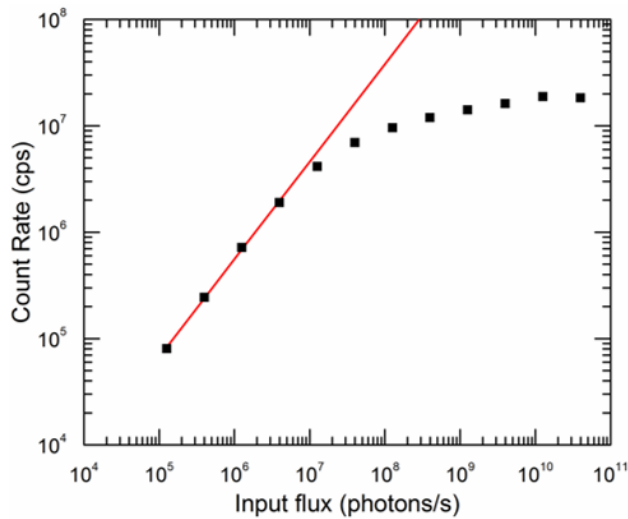

Fig. 6 Measured count rate as a function of the input photon flux (black squares). The red solid line is a linear fit to the first four data points.